\documentclass{aastex63}

\usepackage[finalnew]{trackchanges}

\submitjournal{ApJ}

\shorttitle{CME Radial Expansion From Spacecraft Conjunction}
\shortauthors{Lugaz et al.}

\graphicspath{{./}{figures/}}

\begin{document}

\title{Inconsistencies Between Local and Global Measures of CME Radial Expansion\\ as Revealed by Spacecraft Conjunction\add{s}}

\author[0000-0002-1890-6156]{No\'e Lugaz}
\email{noe.lugaz@unh.edu}
\affiliation{Space Science Center and Department of Physics and Astronomy, University of New Hampshire \\
8 College Rd \\
Durham, NH, 03824, USA}

\author{Tarik M. Salman}
\affiliation{Space Science Center and Department of Physics and Astronomy, University of New Hampshire \\
8 College Rd \\
Durham, NH, 03824, USA}

\author[0000-0002-9276-9487]{R\'eka M. Winslow}
\affiliation{Space Science Center and Department of Physics and Astronomy, University of New Hampshire \\
8 College Rd \\
Durham, NH, 03824, USA}

\author[0000-0002-0973-2027]{Nada Al-Haddad}
\affiliation{Space Science Center and Department of Physics and Astronomy, University of New Hampshire \\
8 College Rd \\
Durham, NH, 03824, USA}

\author[0000-0001-8780-0673]{Charles J. Farrugia}
\affiliation{Space Science Center and Department of Physics and Astronomy, University of New Hampshire \\
8 College Rd \\
Durham, NH, 03824, USA}

\author{Bin Zhuang}
\affiliation{Space Science Center and Department of Physics and Astronomy, University of New Hampshire \\
8 College Rd \\
Durham, NH, 03824, USA}

\author{Antoinette B. Galvin}
\affiliation{Space Science Center and Department of Physics and Astronomy, University of New Hampshire \\
8 College Rd \\
Durham, NH, 03824, USA}

\begin{abstract}
The radial expansion of coronal mass ejections (CMEs) is known to occur from remote observations; from the variation of their properties with radial distance; and from local {\it in situ} plasma measurements showing a decreasing speed profile throughout the magnetic ejecta (ME). However, little is known on how local measurements compare to global measurements of expansion. \remove{and whether radial expansion is mostly driven by the excess pressure in the ME or by the pressure-balanced expansion of the ME within the solar wind}\change{Here, we take advantage of 42 CMEs being measured in the inner heliosphere by two spacecraft in radial conjunction to determine how the magnetic field decreases with distance as a measure of their global expansion}{Here, we present results from the analysis of 42 CMEs measured in the inner heliosphere by two spacecraft in radial conjunction.The magnetic field decrease with distance provides a measure of their global expansion}. \change{As all these CMEs are measured in situ near 1~au by STEREO or Wind}{Near 1 au, the decrease in their bulk speed provides a measure of their local expansion}. We find that these two measures\remove{appear to} have little relation with each other\remove{, even when looking only at the events with the closest conjunctions}. We also \change{determine}{investigate} the relation between characteristics of CME expansion and CME properties. We find that the expansion depends on the initial magnetic field strength inside the ME, but not significantly on the magnetic field inside the ME measured near 1~au. This is an indirect evidence that CME expansion in the innermost heliosphere is driven by the high magnetic pressure inside the ME, while by the time the MEs reach 1~au, they are expanding due to the decrease in the solar wind dynamic pressure with distance. We also determine the evolution of the ME tangential and normal magnetic field components with distance, revealing significant deviations as compared to the expectations from force-free field configurations as well as some evidence that the front half of MEs expand at a faster rate than the back half. 
\end{abstract}


\section{Introduction} \label{sec:intro}

Radial expansion is one of the fundamental characteristics of coronal mass ejections (CMEs), \change{evident both from}{described in early work} using {\it in situ} measurements \citep[]{Klein:1982,Burlaga:1982,Suess:1988}. \add{It is also clearly occurring based on} the fact that CMEs are \add{remotely} imaged as being a fraction of a solar radius wide when they erupt and are on average \add{measured as being} 45 solar radii (0.21 au) when they reach Earth. Associated with this increase in radial size, the magnetic field strength inside the CME decreases \add{as the CME propagates to larger heliocentric distances} \citep[]{Bothmer:1998}. Most of what is known about the increase in radial size and decrease in magnetic field inside magnetic ejecta (MEs) is obtained from statistical studies of {\it in situ} measurements of different MEs at different heliocentric distances. Thus, based on measurements by Helios, ISEE-3, IMP-8, ACE, {\it Wind} and Voyager, using different boundaries and different subsets of CMEs, past studies have found that the radial size of an ME increases as $r^{0.6}$ to $r^{0.9}$ and the magnetic field scales as  $r^{-1.4}$ to $r^{-1.9}$ \citep[]{Bothmer:1998,Liu:2005, Leitner:2007,Gulisano:2010}. This was revisited using STEREO, ACE and MESSENGER data for the solar cycle 24 yielding  almost the same index of radial dependency as $-1.95 \pm 0.19$ \citep[]{Winslow:2015}. This approach provides a measure of the average global expansion and assumes that there is a unique typical behavior of CMEs. Statistical methods would not work well if, for example, fast CMEs always expand differently than slow ones. A different measure of the global CME expansion can be obtained in a case-by-case basis by tracking CME radial size with heliospheric imagers up to distances of about 0.5~au \citep[]{Savani:2009,Nieves:2012,Lugaz:2012b}, which has revealed an expansion on the lower end of the range from statistical studies, as $r^{0.6}$ to $r^{0.8}$. In a recent work, \citet{AlHaddad:2019} compared the index of decrease of the magnetic field with the index of increase of the ME radial size for two different simulations, finding that the initiation mechanism \add{and the CME propagation speed} do not appear to have a large influence on the ME expansion in the innermost heliosphere.

\begin{figure}[tb]
\centering
{\includegraphics*[width=.98\linewidth]{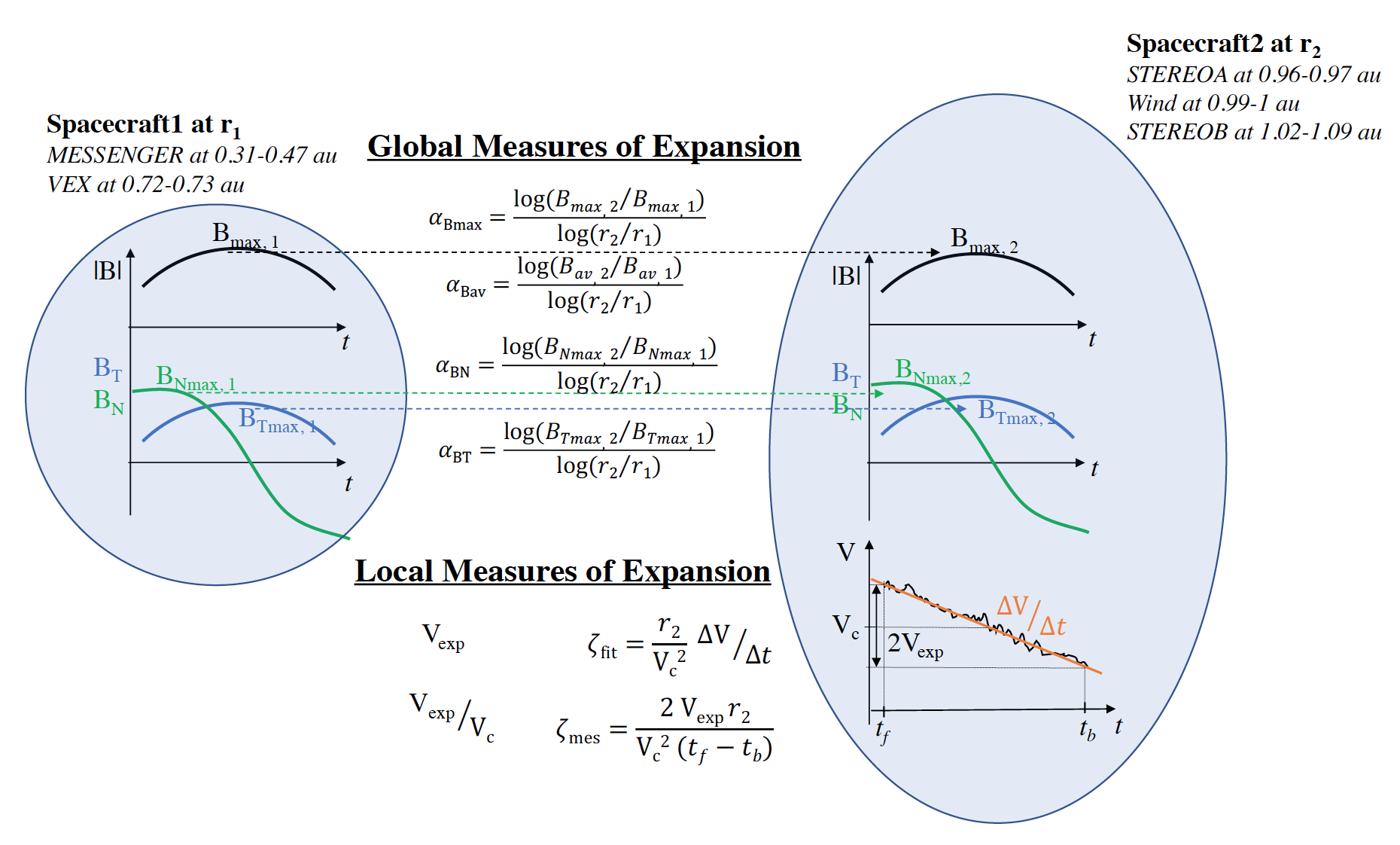}}
\caption{Schematic representation and definitions of the global and local measures of ME expansion. The idealized ME cross-section and associated magnetic field measurements are shown at two locations at different heliocentric distances. Comparing measurements from these two locations define the global expansion. At the second spacecraft, measurements of the plasma velocity allow to derive various measures of the local expansion.}
\label{fig:sketch}
\end{figure}

Another measure of CME expansion can be obtained from the direct analysis of {\it in situ} measurements at a given location, as the large majority of MEs have a decreasing speed profile. This is clearly a local measure. Figure~\ref{fig:sketch} shows schematic representations of the various measures of CME expansion. The expansion speed, defined as half the front-to-back speed difference is found to vary from a few tens of km\,s$^{-1}$ to as much as 250 km\,s$^{-1}$  \citep[]{Burlaga:1982,Farrugia:1993}. \citet{Klein:1982} noted that the expansion speed is on the order of half the ambient Alfv{\'e}n speed, meaning that expansion occurs sub-Aflv{\'e}nically. \citet{Gosling:1994} and \citet{Reisenfeld:2003} presented the observations of several CMEs which were bounded by a forward-reverse shock pair. This shock pair was attributed to the CME expansion becoming super-fast due to high pressure inside the ME. This type of over-expanding structure has only been reported away from the ecliptic with Ulysses observations (at latitudes greater than 22$^\circ$). In a recent study, \citet{Lugaz:2017b} showed that slow CMEs may drive shocks because of their radial expansion in the ecliptic plane, although the expansion remains sub-Alfv{\'e}nic. Such shocks may form at distances of 0.2 au or greater, depending on the rate at which the CME expansion speed and Alfv{\'e}n speed decrease \citep[]{Poedts:2016,Lugaz:2017}. 

A difficulty with studying CME expansion is that the expansion speed is found to depend significantly on the CME size and propagation speed, with larger and faster CMEs having larger expansion speeds \citep[]{Owens:2005,Gulisano:2010}. To solve this problem, researchers have focused on a dimensionless expansion parameter, typically the ratio of the expansion to propagation speed. \citet{Demoulin:2009} and \citet{Gulisano:2010} developed a different formalism, in which a dimensionless expansion parameter, $\zeta$, is defined as follows: 

\begin{equation}\zeta =  \frac{D}{V_c^2} \frac{\Delta V}{\Delta t} \sim \frac{D}{S} \frac{2 V_\mathrm{exp}}{V_c}.\end{equation}

Here, $D$ is the heliospheric distance where the measurements are made, $S = V_c \Delta t$ is the CME size, $V_\mathrm{exp}$ and $V_c$ are the CME expansion and center speeds, respectively, $\frac{\Delta V}{\Delta t}$ is the slope of the CME velocity time profile. This dimensionless parameter scales as $V_c^{-2}$, taking into consideration that faster and wider CMEs have higher expansion speed. Based on measurements in the inner heliosphere for several dozen isolated CMEs, the authors found that $\zeta$  clusters around 0.8 \citep[]{Demoulin:2010b}. From a theoretical analysis, \citet{Gulisano:2010} argued that this local measure should represent the global expansion of CMEs with the CME size growing as r$^\zeta$ and the magnetic field strength decreasing as r$^{-2\zeta}$. Note that the formula uses the slope of the velocity profile $\frac{\Delta V}{\Delta t}$, which is equivalent to using the expansion speed only for those cases where the velocity can be fitted linearly for the entire ME duration. 

The physical cause of CME expansion is still a matter of debate, although it is generally agreed that it is associated with pressure balance or imbalance between the ME and the solar wind. It has been proposed that CME expansion is associated with the pressure imbalance between the high pressure of the magnetically dominated ME and the lower pressure in the solar wind \citep[]{Klein:1982}. In that sense, CME expansion is associated with over-pressure. A somewhat different explanation is that CME expansion is related to the pressure balance between the ME and the solar wind, {\it i.e} between the ME magnetic pressure and the solar wind dynamic pressure. The fact that the solar wind pressure decreases with heliospheric distance then implies that CMEs keep on expanding as they propagate outward \citep[]{Demoulin:2009,Gulisano:2010}. Lastly, \citet{Suess:1988} argued that measurements of decreasing speed profile inside MEs are associated with magnetic tension and the necessary plasma motion to maintain a force-free state of the ME. 

Very few studies have investigated CME size or expansion from multiple {\it in situ} measurements in near-conjunction for more than one CME event. The exceptions are the study of \citet{Leitner:2007}, which focused on 7 CMEs measured in conjunction (4 with measurements below 1~au), the recent study by \citet{Good:2019}, which focuses on 18 events and the study by \citet{Vrsnak:2019}, which focuses on 11 events during the cruise phase of MESSENGER and VEX. In particular, \citet{Good:2019} found a significant difference between the power-law obtained from performing a fit of the maximum magnetic field with distance ($-1.76 \pm 0.04$) as compared to the average of the power-law indices of these 18 events ($-1.34 \pm 0.71$). The same type of results was obtained for the larger statistics of \citet{Salman:2020}. Here, we further dive into these datasets to investigate ME expansion. \add{We note that near-conjunction is often taken quite loosely, as has also been done here. Angular separations for spacecraft considered in near-conjunction in these studies typically range from 1-20$^\circ$ with a few cases up to 30$^\circ$ with the average angular separation being $\sim 5^\circ$ in the study of} \citet{Good:2019} \add{and $\sim 16^\circ$ for the dataset of} \citet{Salman:2020}. 

To learn more about CME expansion, it is essential to compare its local measures (the $\zeta$ parameter, the expansion speed, etc.) with global ones (how much do the CME size and magnetic field change with distance). For example, the use of the dimensionless index of \citet{Demoulin:2009} is meant to take into consideration the fact that fast and wide CMEs may have a large front-to-back speed difference without having a large expansion {\it per se}. However, this begs the question of the cause of the large size of these CMEs. Is it related to their expansion earlier on or a large size near the Sun? Performing such a study has not been possible until now because it requires the investigation of CME expansion in both its global and local ways in a case-by-case basis for enough events to compare with past statistical studies. Here, we take advantage of the numerous CME events measured in conjunction between two spacecraft in the inner heliosphere as recently presented in \citet{Salman:2020} using data from MESSENGER, Venus Express (VEX), {\it Wind} and STEREO. In section~\ref{sec:data}, we quickly summarize our data and procedure. In section~\ref{sec:results}, we compare the different measures of CME expansion with each other and with other related CME properties. In section~\ref{sec:discuss}, we discuss and conclude.

\section{Data and Methods} \label{sec:data}
\citet{Salman:2020} presented 47 two-spacecraft conjunction measurements of CMEs over the first half of solar cycle 24, from 2008 to 2014 for spacecraft longitudinal separations of less than 35$^\circ$, with 8 events measured at less than 5$^\circ$ separations and 20 at less than 15$^\circ$ separations. Five events were conjunction between Venus Express (VEX) and MESSENGER, 18 conjunction events occurred between MESSENGER and a spacecraft near 1 au ({\it Wind}, STEREO-A or STEREO-B), and 24 between VEX and a spacecraft near 1~au. Since STEREO and {\it Wind} have plasma instruments, we have {\it in situ} measurements of the CME speed near 1~au for these 42 CMEs in addition to magnetic field measurements at two different distances. For the five conjunctions events between MESSENGER and VEX, we do not have any plasma measurements. Our analysis thereafter focuses on the 42 events with plasma measurements near 1~au.  Because Mercury's heliocentric distance (and therefore MESSENGER's) varies between 0.31 and 0.47 au, whereas Venus stays at 0.72-0.73~au, we have measurements over distances varying from a factor of 1.3 (Venus to STEREO-A) to a factor of  3.2 (Mercury at perihelion to STEREO-B). 

The magnetic field decrease with heliospheric distance for this dataset is presented by \citet{Salman:2020} who found a decrease of the maximum field, $B_{\max}$, inside the ME with an index of $-1.91 \pm 0.25$. Although most events have gaps in measurements corresponding to the time when MESSENGER or VEX are inside their planetary magnetosphere, we can do the same study with the average magnetic field, $B_{av}$ for which we find an index of $ -1.87 \pm 0.32$ excluding the 5 MESSENGER-VEX conjunctions. For each conjunction event, we also calculate the quantity $\alpha_B$ \citep[see for example][]{Dumbovic:2018}:
$$\alpha_B = \frac{\log \left( {B_2/B_1} \right) }{\log \left( {r_2/r_1} \right) },$$
where indices 1 and 2 correspond to the first \add{(closer to the Sun)} and second \add{(further away from the Sun)} spacecraft, respectively. We do so for both the maximum magnetic field ($\alpha_{Bmax}$), the average magnetic field  ($\alpha_{Bav}$), as well as the maximum value of the tangential ($T$) and normal ($N$) magnetic field components inside the ME ($\alpha_{BT}$ and $\alpha_{BN}$). 

In addition, near 1~au, we derive local measures of the ME radial expansion: 1) the dimensionless expansion parameter $\zeta_{fit}$ using the procedure of \citet{Gulisano:2010}, {\it i.e.} by performing a linear fit on the velocity data to derive $\Delta V /\Delta t$. We also calculate 2) the expansion speed $V_\mathrm{exp}$, 3) the ratio of the expansion speed to the CME speed $V_\mathrm{exp}/V_\mathrm{center}$, and 4) $\zeta_{mes}$ using the measured value of $\Delta V = 2 V_\mathrm{exp}$ in equation (1) rather than the fit to the velocity data. We also use or derive associated CME properties: its initial speed from coronagraph \citep[as listed in][]{Salman:2020}, its size near 1 AU (using the average CME speed), and the maximum and average magnetic field inside the ME. 

\begin{figure}[tb]
\centering
{\includegraphics*[width=.58\linewidth]{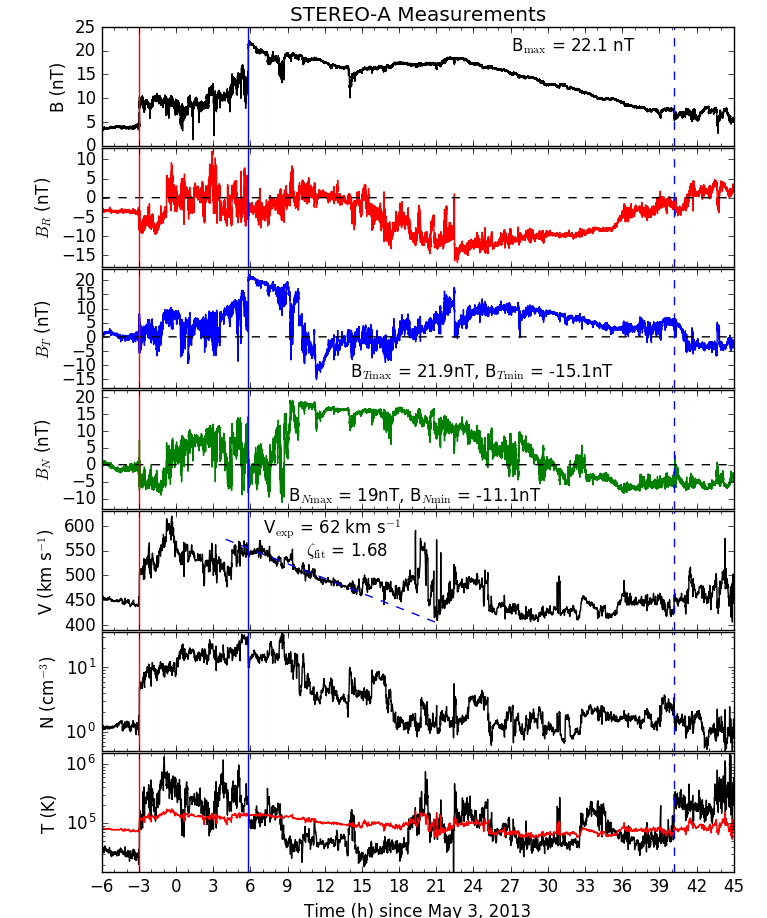}}
{\includegraphics*[width=.4\linewidth]{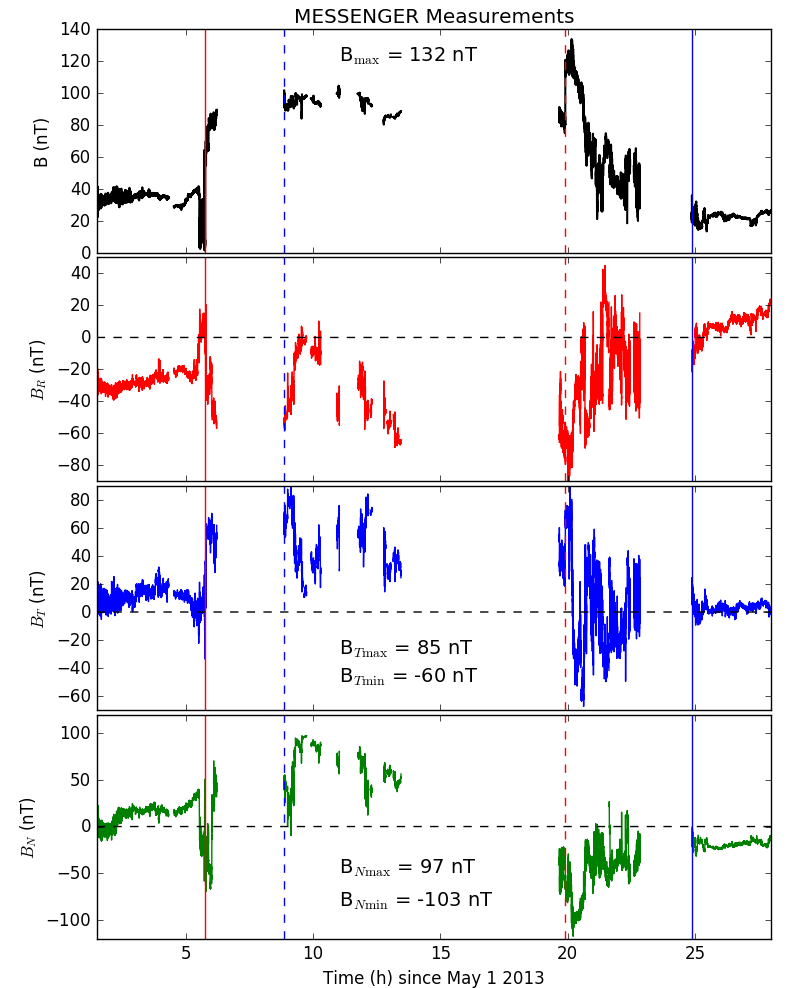}}
\caption{2013 May 1--4 CME measured at STEREO-A ({\it left}) and MESSENGER ({\it right}). A linear trend in the velocity can be found in about the first 30\% of the ME at STEREO-A. This is used to derive $\zeta_\mathrm{fit}$. The maximum and average of the magnetic field magnitude as well as the maximum and minimum of $B_T$ and $B_N$ are used to derive various exponent decrease $\alpha$. The red line marks the shock arrival at STEREO-A and MESSENGER, the blue lines mark the ME boundaries with dashed lines used when the boundary's location is not certain. For the MESSENGER data, the dashed red line shows a shock propagating inside the ME.}
\label{fig:example}
\end{figure}

\section{Results: Specific Events} \label{sec:results_spec}
Most of the best conjunction events have been studied in detail in previous work. Here, we present one additional event to illustrate our technique and summarize results for three previously published events. The four events we highlight are among the eight best conjunctions (separations of less than 5$^\circ$) with data near 1~au. The results of the analysis described below for these four events are listed in Table~\ref{tab:example}. 

\subsection{2013 May 1--4 CME: MESSENGER-STEREOA conjunction}
The May 1--4, 2013 CME event \citep[event 14-2013 in][]{Salman:2020} is a conjunction between MESSENGER (at 0.36~au) and STEREO-A (at 0.96 au) when the longitudinal separation between the two spacecraft was only $\sim 2.9^\circ$. The measurements at MESSENGER and STEREO-A are shown in Figure~\ref{fig:example}.  This was a moderately fast event with a coronagraphic speed of 700~km\,s$^{-1}$ and a maximum ME speed near 1~au of 570~km\,s$^{-1}$. \add{In most cases, solar wind plasma measurements are not available at Mercury with MESSENGER}. The maximum ME magnetic field of 132 nT at MESSENGER and 22 nT at STEREO-A results in a value of $\alpha_{Bmax} = -1.84$, which is relatively typical. The exponent for the average magnetic field is very similar at $\alpha_{Bav} = -1.89$. The speed profile at 1~au is complex and we consider that a linear trend in the velocity can only be found for the front 30\% of the ME. Using this limited period, a linear fit to the velocity profile implies that $\zeta_{fit}$ = 1.7. The expansion speed as measured from maximum to minimum is 62~km\,s$^{-1}$ and the ME center speed is about 485~km\,s$^{-1}$. Using the measured $V_\mathrm{exp}$, we can derive $\zeta_{mes} = 0.61$. The ratio of expansion to center speeds is $\sim 13$\%. From this event, we can already see a disagreement between the global decrease in the magnetic field strength and the local measure, especially using the procedure of \citet{Gulisano:2010}. 

However, one can also see that the peak in the magnetic field at MESSENGER occurs in what appears to be a discontinuity or fast forward shock near the back of the ME, whereas there is no such signature at Earth. Excluding this period, we find that $B_{\max} = 102$~nT and $\alpha_B = -1.57$. The presence of fast-forward shocks at the back of MEs was discussed in \citet{Lugaz:2015a}. \add{Such a shock/discontinuity was not observed at STEREO-A. Based on past work, this raises two possibilities: i) the shock fully propagated through the ME before the ME impacting STEREO-A. In that case, the period of compression by the shock is expected to be followed by a period of over-expansion} \citep[]{Gulisano:2010,Lugaz:2012b}. \add{Depending on the timing of this exit, the ME global and local measures of expansion may be affected. ii) The shock dissipated as it propagated inside the ME} \citep[]{Farrugia:2004,Lugaz:2007} \add{and only the back half got affected}. \add{In both cases, the back half of the ME may have been compressed, resulting in the flat velocity profile in the back measured near 1~au. At MESSENGER, there is no clear driver for this shock as the magnetic field strength goes back to normal values a few hours after the shock. At such, it is unlikely that the ME measured near 1~au is the result of the merging of two CMEs}.

The decrease of the tangential and normal magnetic field components for the front of the ME (positive values) is $\alpha_\mathrm{Tfront} = -1.40$ and $\alpha_\mathrm{Nfront} = -1.68$. The back (negative values) for which the peak occurs after the shock at Mercury are $\alpha_\mathrm{Tback} = -1.41$ and $\alpha_\mathrm{Nback} = -2.29$. This shows that the normal (north-south) component of the magnetic field decreased a bit faster than the tangential (east-west) component, but also highlights how this detailed analysis may be affected by the presence of shocks and  ``datagaps'' in MESSENGER measurements associated with magnetospheric crossings. 

\subsection{Other Events}

\citet{Good:2015} and \citet{Salman:2020} presented a different conjunction that occurred on November 4--8, 2011 (event 8-2011) with an initial speed of 750~km\,s$^{-1}$  and a maximum ME speed near 1~AU of 440~km\,s$^{-1}$. Although a different section of the same event also impacted Venus, the best conjunction is between MESSENGER and STEREO-B ($\sim 4.8^\circ$ longitudinal separation). For this event, $\alpha_{Bmax} = -1.93$ and $\alpha_{Bav} = -1.80$, but there is a large expansion speed at 1 AU of $\sim 85$~km\,$^{-1}$ 
\begin{table}[h]
\caption{Examples from past studies and Figure~\ref{fig:example}. Results with a $^*$ indicate cases for which the peak is likely to have occurred during a magnetospheric path of MESSENGER and is therefore likely missed. Values in parentheses for $B_T$ or $B_N$ correspond to the decrease for that component of the magnetic field in the back half of the ME.}
\centering
\small
\addtolength{\tabcolsep}{-2pt}
\begin{tabular}{|c|c|c|c|c|c|c|c|c|c|}
\hline
Event & Sep.\ & $V_\mathrm{init}$ & $\alpha_{Bmax}$ & $\alpha_{Bav}$ & $\alpha_{BT}$ & $\alpha_{BN}$ & $\zeta_\mathrm{fit}$ & $\zeta_\mathrm{mes}$ & $V_\mathrm{exp}/V_\mathrm{c}$\\
\hline
8-2011 & 4.8$^\circ$ & 950\,km\,s$^{-1}$& $-1.9$ & $-1.8$ & $-1.8$ ($-1.5$) & $-2.4$ ($-1.5$)& 1.5 & 0.95 & 0.18 \\
9-2011 & 4.6$^\circ$ & 760\,km\,s$^{-1}$& $-2.0$ & $-1.6$ & $-2.0$ ($-2.1$) & $-1.8$ ($-2.2$)& 0.19 & 0.15 & 0.01 \\
14-2013 & 2.9$^\circ$ & 700\,km\,s$^{-1}$& $-1.8$ & $-1.9$ & $-1.4$ ($-1.4$)& $-1.7$ ($-2.3$) & 1.7 & 0.61 & 0.13\\
21-2013 & 3.1$^\circ$ & 700\,km\,s$^{-1}$& $-1.6$ & $-1.4$ & $-1.2^*$ & $-1.4$ ($-1.1$) & 0.67 & 0.51 & 0.10\\
\hline
\end{tabular}
\label{tab:example}
\end{table}
corresponding to $\zeta_{fit} = 1.5$, $\zeta_{mes} = 0.95$ and a ratio of the expansion to the center ME speed of 18\%. Once again, local and global measures of expansion disagree.

\citet{Winslow:2016} presented a complex  conjunction event (event 9-2011) between MESSENGER (at 0.42~au) and STEREO-A ($\sim 4.6^\circ$ longitudinal separation) on December 30, 2011 -- January 1, 2012 with an initial speed of 950~km\,s$^{-1}$  and a maximum ME speed near 1~au of 630~km\,s$^{-1}$. For this event, $\alpha_{Bmax} = -1.99$ and  $\alpha_{Bav} = -1.63$, whereas measurements near 1~au show a nearly flat velocity profile with an expansion speed of only 10~km\,s$^{-1}$, $\zeta_{fit} = 0.19$ and $\zeta_{mes} = 0.15$. In this case, the decrease of the magnetic field inside the ME with distance is typical, but the bulk speed profile at 1 AU indicates a lack of expansion near 1 AU. This case is somewhat unusual because of the complex interaction with the heliospheric current sheet that is found to be engulfed inside the ME at 1 AU. The ratio of expansion to center speeds is of the order of 1\%. 

Another event (event 21-2013) was recently discussed in \citet{Lugaz:2020} for a conjunction between MESSENGER and L1 ($\sim 3.1^\circ$ longitudinal separation) on July 11--14, 2013 with an initial speed of 600~km\,s$^{-1}$  and a maximum ME speed near 1~au of 500~km\,s$^{-1}$. 
\begin{table}[h]
\caption{Average values and 1-$\sigma$ standard deviations obtained in this study and comparison to past studies. The first four quantities are obtained in our study by measuring the magnetic field at two spacecraft in conjunction, while for the past studies, these are typically from fits to different MEs measured at different heliocentric distances. The seven other quantities are obtained from measurements near 1~au.  F05: \citet{Farrugia:2005}, L07: \citet{Leitner:2007}, W15: \citet{Winslow:2015}, G19: \citet{Good:2019}, W05: \citet{WangC:2005}, L05: \citet{Liu:2005}, G10: \citet{Gulisano:2010}, D08: \citet{Demoulin:2008}, RC10: \citet{Richardson:2010}, J18: \citet{Jian:2018}, NC18: \citet{Nieves:2018} and L18: \citet{Lepping:2018}. The data source in the inner heliosphere is indicated in parentheses (H: Helios, P: Pioneer Venus Orbiter, M: MESSENGER, V: {\it Venus Express}). For $\alpha_{Bmax}$, we list two values for G19, the first one using the same procedure as done here but for 13 events and the second one using a fitting procedure.}
\centering
\small
\addtolength{\tabcolsep}{-2pt}
\begin{tabular}{|c|c|c|c|c|}
\hline
Quantity & Average $\pm$ $\sigma$ & Past Results & Source\\
\hline
$\alpha_{Bmax}$ & $-1.81$ $\pm$ 0.84  & -1.73, -1.64 $\pm$ 0.40, -1.89 $\pm$ 0.14(3-$\sigma$), &  F05(H), L07(H,P), W15(M)\\
& & -1.34 $\pm$ 0.71, -1.76 $\pm$ 0.04(3-$\sigma$) & G19(M,V)\\
\hline
$\alpha_{Bav}$ & $-1.91$ $\pm$ 0.85 & -1.38, -1.52, -1.4 $\pm$ 0.08, & F05(H), W05(H,P), L05(H)\\
 & &  -1.85 $\pm$ 0.07,  -1.95 $\pm$ 0.19(3-$\sigma$)  &  G10(H), W15(M) \\
 \hline
$\alpha_{BT}$ & $-1.71$ $\pm$ 0.67 & & \\
\hline
$\alpha_{BN}$ & $-1.76$ $\pm$ 0.65 & & \\
\hline
\hline
$\zeta_\mathrm{fit}$ & 0.95 $\pm$ 1.05 & 0.81 $\pm$ 0.19, 0.7 $\pm$ 0.61 & D08, G10(H) \\
\hline
$\zeta_\mathrm{mes}$ & 0.43 $\pm$ 0.52 & 0.45 & RC10 \\
\hline
$V_\mathrm{exp}$ (km\,s$^{-1}$) & 32 $\pm$ 42 &  31 $\pm$ 3, 62 $\pm$ 3, 28 & RC10, J18, NC18 \\
\hline
$V_\mathrm{center}$ (km\,s$^{-1}$) & 449 $\pm$ 131 &  476 $\pm$ 6, 445, 434, 436 & RC10, J18, NC18, L18 \\
\hline
$V_\mathrm{exp}/V_\mathrm{center}$ &  0.066 $\pm$ 0.085 & & \\
\hline
$S_\mathrm{ME}$ (AU) & 0.29 $\pm$ 0.14 & 0.33 $\pm$ 0.01, 0.22 $\pm$ 0.11 & RC10, L18 \\
\hline
$<B_\mathrm{ME}>$ (nT) & 10.8 $\pm$ 4.4 & 10.1 $\pm$ 0.3, 11 & RC10, NC18 \\
\hline
\end{tabular}
\label{tab:stats}
\end{table}
This long-duration event is found to have $\alpha_{Bmax} = -1.58$ and $\alpha_{Bav} = -1.38$ and $\zeta_{fit} = 0.67$ for an expansion speed of about 50~km\,s$^{-1}$ corresponding to $\zeta_{mes} = 0.51$ and $V_\mathrm{exp}/V_\mathrm{center} \sim 10$\%. 

With the exception of the case highlighted in \citet{Lugaz:2020}, these four examples highlight the following: even for conjunctions between two spacecraft with longitudinal separations of less than 5$^\circ$, the local and global measures of expansion do not necessarily agree. In the following section, we look at overall results for all CMEs and compare the various measures of expansion with each others.

\section{Results: Statistics}\label{sec:results}
\subsection{Average Values}

Table~\ref{tab:stats} shows the statistics of the values of the different measures of CME expansion and other CME properties, as well as comparison to previous studies in the inner heliosphere (excluding other studies with Ulysses or Voyager data past 2~au). Throughout, we give the 1-$\sigma$ standard-deviation as an error bar when quoting a value. 

For $\alpha_{B\max}$, the average is $-1.81 \pm 0.84$ for the full dataset but $-1.67 \pm 0.43$ for the conjunctions between MESSENGER and 1~au and $-1.93 \pm 1.06$ for the conjunctions between VEX and 1~au. In all cases, this is a similar average as compared to the results from statistical studies but with a much larger standard-deviation, as for example \citet{Winslow:2015} found a 3-$\sigma$ value of $\pm 0.19$. Combining all past studies, a range for the exponent decrease of $B_{\max}$ can be obtained as $-1.75 \pm 0.4$. Only 20 out of the 42 events studied here are within this range. For $\alpha_{Bav}$, we find an average of $-1.96 \pm 0.90$. 
The average value for $\zeta$ is 0.86 $\pm 0.83$, comparable to past studies. We note that we are able to identify a linear trend in the velocity profile for 82\% $\pm$ 22\% of the ME duration (for 31 events, a trend is identified for more than 60\% of the ME duration). In fact, the event highlighted in Figure~\ref{fig:example} is the one for which the linear trend is the least clear. The average expansion speed is 32 $\pm$ 44~km\,s$^{-1}$ (or 39 $\pm$ 35~km\,s$^{-1}$ if excluding three contracting events) and that of the dimensionless expansion is 0.067 $\pm 0.09$ (0.083 $\pm 0.06$ if excluding these three events). The average size of the MEs near 1~au is 0.29 $\pm 0.14$~au, comparable to that found from all {\it ACE} MEs by \citet{Richardson:2010}. That number is larger than the canonical 0.21~au from \citet{Lepping:2018} but the latter is obtained for a force-free fit to the data, whereas our number and that by  \citet{Richardson:2010} are simply derived by integrating the solar wind speed with time during the ME passage. For all quantities, the average values are within the typical ranges from past studies, highlighting that our dataset is not biased.

\begin{figure}[tb]
\centering
{\includegraphics*[width=.48\linewidth]{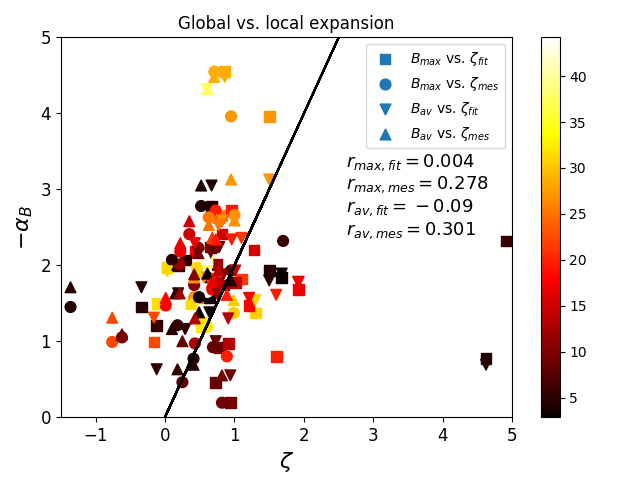}}
{\includegraphics*[width=.48\linewidth]{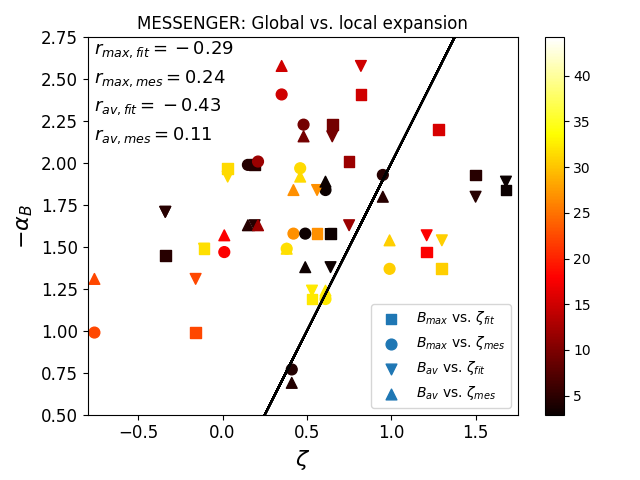}}
{\includegraphics*[width=.48\linewidth]{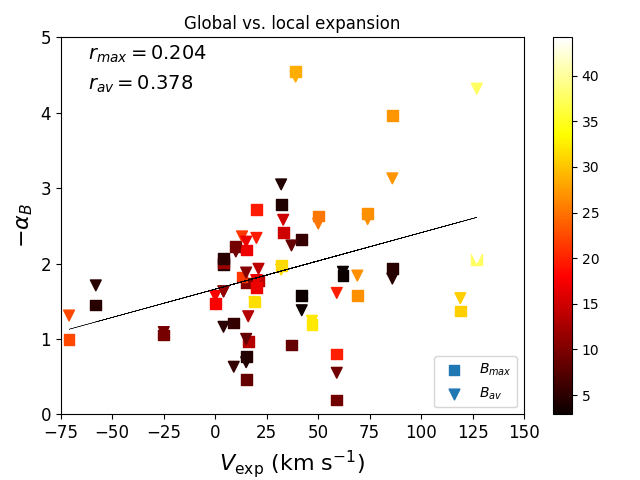}}
{\includegraphics*[width=.48\linewidth]{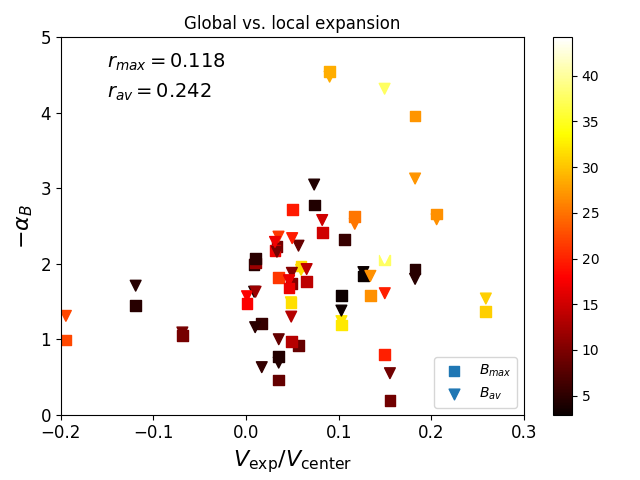}}
\caption{Global ($y$-axis) vs.\ local ($x$-axis) measures of CME expansion. The top panels show the index decrease of the magnetic field, $\alpha$, as compared to the dimensionless expansion parameter near 1~AU, $\zeta$. The line show the expected value of $\alpha = -2\zeta$. The bottom panels show $\alpha$ as compared to the ME expansion speed near 1~AU (left) and the ratio of ME expansion to  center speeds (right). The thin line shows the linear relation for the best fit: $-\alpha_{Bav} =  1.66 + 0.0075 V_\mathrm{exp}$. All data points are color-coded with the angular separation between the two spacecraft with the scale in $^\circ$ given on the right-hand colorbar.}
\label{fig:stats_zeta}
\end{figure}

\subsection{Comparison of Local and Global Measures of CME Expansion}

In the top panels of Figure~\ref{fig:stats_zeta}, we show plots of $\alpha_{Bmax}$ and $\alpha_{Bav}$ as compared to $\zeta_\mathrm{fit}$ and $\zeta_\mathrm{mes}$ as well as the values when the first spacecraft is MESSENGER rather than VEX since the $\alpha$ values have less variability when the former is the first spacecraft rather than the latter. The symbols are color-coded with the spacecraft angular separation and the top panels show the line $\alpha = 2 \zeta$, which is the expected trend. The data is un-correlated for $\zeta_\mathrm{fit}$ (obtained by fitting to the slope of the velocity), while there is a very weak correlation with $\zeta_\mathrm{mes}$ (calculated using the measured expansion speed) with the highest correlation coefficient, $r =  0.3$ for the average magnetic field. 

In the bottom panels of Figure~\ref{fig:stats_zeta}, $\alpha_{Bmax}$ and $\alpha_{Bav}$ are compared to $V_\mathrm{exp}$ and $V_\mathrm{exp}/V_\mathrm{center}$. This again compares global quantities of CME expansion (in the $y$-axis) to local quantities near 1~au of the CME expansion (in the $x$-axis). The largest correlation coefficient is found between the $\alpha$ index for the average magnetic field and the expansion speed and is $r = 0.378$. Other correlation coefficients are below 0.25. From these plots and the correlation values, it is clear that, irrespective of the exact quantities being compared, local and global measures of CME expansion are at best weakly related. In particular, even for the smaller angular separations and large radial separations (the MESSENGER plot), small and large values of $\alpha$ are associated with typical values of $\zeta$ around its average of 0.7.

\subsection{Correlation of Global Measures of CME Expansion with Other CME Properties}

\begin{figure}[tb]
\centering
{\includegraphics*[width=.48\linewidth]{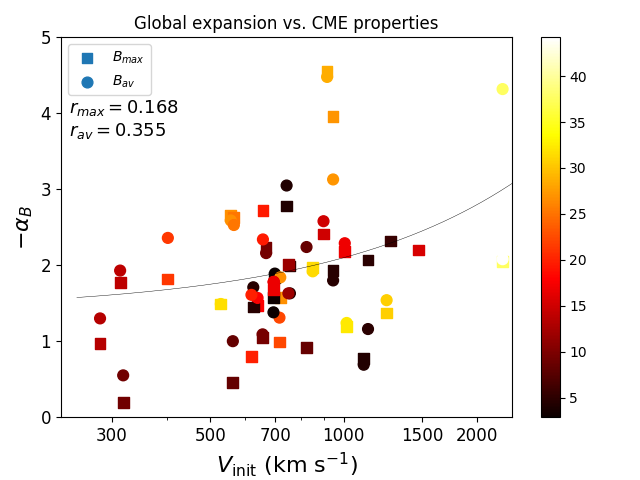}}
{\includegraphics*[width=.48\linewidth]{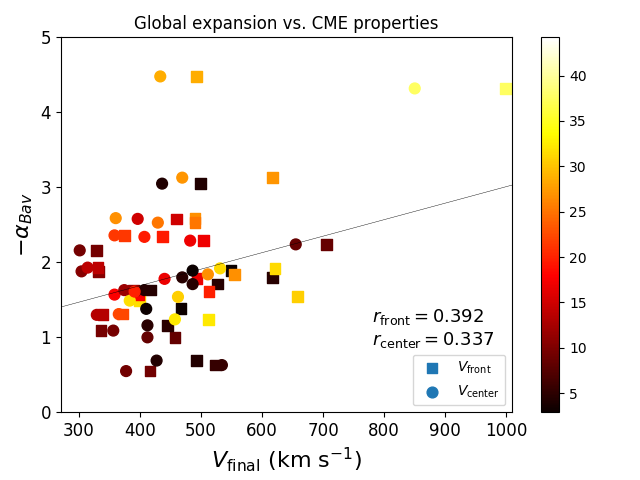}}
\caption{Global measures of CME expansion ($y$-axis) vs.\ CME speed. The left panel shows the index decrease of the magnetic field, $\alpha$, as compared to the initial plane-of-sky coronagraphic speed. The thin line shows the linear relation for the best fit (in a log-linear plot): $-\alpha_{Bav} =  1.40 + 0.0007 V_\mathrm{init}$. The right panel shows $\alpha_{Bav}$ as compared to the final front and center ME speeds near 1~AU. The thin line shows the linear relation for the best fit: $-\alpha_{Bav} =  0.81 + 0.0022 V_\mathrm{front}$. All data points are color-coded with the angular separation between the two spacecraft with the scale in $^\circ$ given on the right-hand colorbar.}
\label{fig:stats_V}
\end{figure}

\begin{figure}[tb]
\centering
{\includegraphics*[width=.48\linewidth]{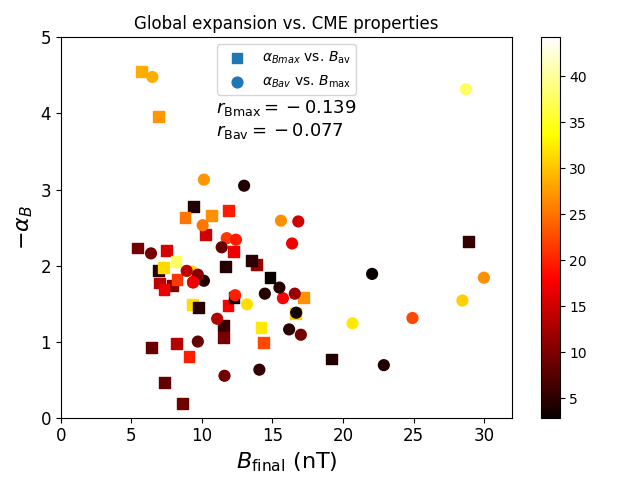}}
{\includegraphics*[width=.48\linewidth]{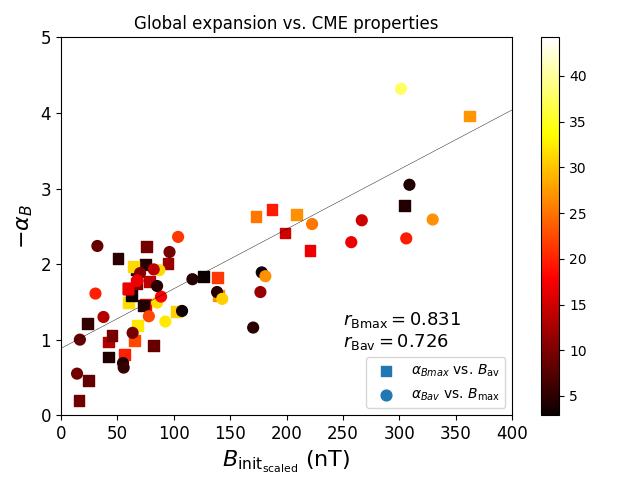}}
{\includegraphics*[width=.48\linewidth]{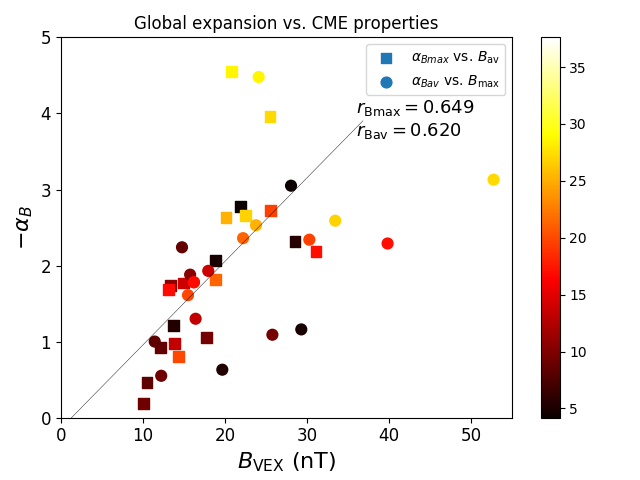}}
{\includegraphics*[width=.48\linewidth]{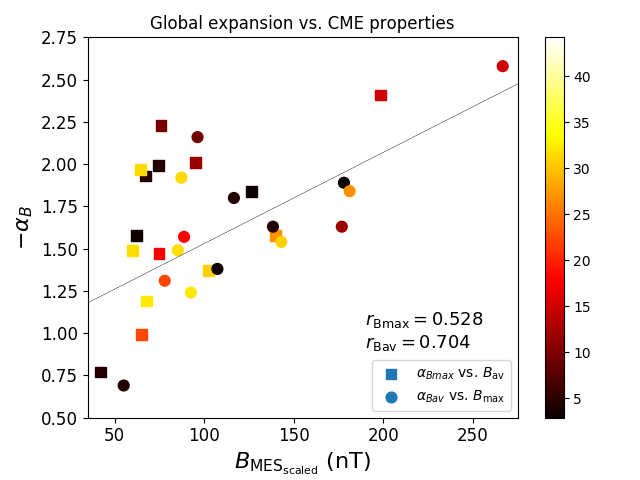}}
\caption{Global measures of CME expansion ($y$-axis) vs.\ ME magnetic fields measured or scaled to various distances. The top left panel shows the index decrease of the magnetic field, $\alpha$, as compared to the ME magnetic field strength measured near 1~AU. The top right panel shows $\alpha$ as compared to the ME magnetic field strength measured by the spacecraft closest to the Sun (VEX or MESSENGER) and scaled to 0.308~au (see text for details). The bottom panels show $\alpha$ as compared to the ME magnetic field measured by VEX (left) and measured by MESSENGER and scaled to 0.308~au (right). The colorbars are the same as in Figures 3 and 4.}
\label{fig:stats_B}
\end{figure}

We extend the analysis of local and global measures of expansion to determine whether other CME properties are correlated with CME global expansion. We focus on the CME initial speed, obtained from the best-observing coronagraph, as explained in \citet{Salman:2020}, the CME final speed measured near 1 AU, as well as the CME magnetic field strength as measured at various distances. Figures~\ref{fig:stats_V} and \ref{fig:stats_B} show the results for $\alpha$ as compared to the ME velocity and magnetic field, respectively, in the same format as Figure~\ref{fig:stats_zeta}.

The ME expansion is only weakly correlated with the CME initial speed, with faster CMEs expanding more rapidly in the inner heliosphere. This correlation remains present near 1~au for $\alpha_{Bav}$ as compared to the CME front and center speeds. It is only a weak correlation but reflects that faster CMEs do expand more strongly in a statistical sense.  We note that the dimensionless analysis of \citet{Dasso:2009} and \citet{Gulisano:2010} results in $\zeta$ being approximately independent of the CME speed, but here we find a weak correlation between $\alpha_\mathrm{av}$ and the CME speed. \add{It is possible that a stronger correlation would exist if the speed was measured at the first spacecraft or if the expansion was calculated for distances closer to the Sun, where expansion may be more related to the initial characteristics of the CME}.   

We then direct our attention to the correlation between $\alpha$ and the magnetic field inside the ME at various distances.
When we compare the $\alpha$ parameter with magnetic field measurements, we only compute the correlation of $\alpha_{Bav}$ with $B_\mathrm{max}$ and of $\alpha_{Bmax}$ with $B_\mathrm{av}$. This way, the values of the magnetic field used to calculate $\alpha$ are not compared with the same values measured at various locations. We note however, that $B_\mathrm{max}$ and $B_\mathrm{av}$ are obviously very well correlated (correlation coefficient $\sim$ 0.81 near 0.72 au and near 1~au), so this may affect the results.

As shown in the top left panel of Figure~\ref{fig:stats_B}, we find no correlation between $\alpha$ with the magnetic field (average or maximum) measured near 1~au with a correlation coefficient below 0.15, whether or not it is corrected for the difference in heliocentric distance between the various spacecraft (see below for details). However, we find a much stronger correlation with the magnetic field measured by Venus Express, with a correlation coefficient of 0.62-0.65 (bottom left panel). 

Correlating the magnetic field measured by MESSENGER with the CME expansion is not straight-forward, because MESSENGER heliocentric distance in our sample varies between 0.308 and 0.466~au. For a typical decrease of the magnetic field as $r^{-1.75}$, this means that the magnetic field would decrease by more than a factor of 2 between these two distances. In comparison, VEX is always between 0.72 and 0.73~au and the variation in magnetic field strength between STEREO-A at 0.96~au and STEREO-B at 1.09~au is only by a factor of 1.25. To correct for the variation in the heliocentric distance of MESSENGER, we scale all measurements to 0.308~au (the measurement made at the lowest heliocentric distance) using the $\alpha$ value obtained for this particular CME. The results show a strong correlation (bottom right panel of Figure~\ref{fig:stats_B}). It should be noted that we use (for example) the value of $\alpha_{Bav}$ obtained for a particular CME to scale the value of $B_\mathrm{av}$ measured for this CME by MESSENGER to 0.308~au and compare it with $\alpha_{Bmax}$. As such, we use fully separated measurements to determine the correlation. Lastly, we scale all VEX and MESSENGER measurements to 0.308~au and obtain very significant correlations between the scaled value of $B$ in the inner heliosphere and $\alpha$, the expansion index (top right panel of Figure~\ref{fig:stats_B}). 

We interpret these results as follows: in the innermost heliosphere, there is clear positive correlation between the ME maximum magnetic field and the expansion index, {\it i.e.},  that MEs with higher internal magnetic pressure in the innermost heliosphere expand more on their way to 1~au. However, near 1~au, there is no relation between the internal magnetic pressure and how much expansion occurred. 

In addition, the range of ME average magnetic fields  is narrower near 1~au than near 0.72~au (at VEX) and at MESSENGER. In our sample, the average ME magnetic field at 1~au is 10.3~nT~$\pm~33\%$ (with STEREO measurements scaled to 1~au), at VEX, it is 18.6~nT~$\pm~36\%$ and at MESSENGER it is 87~nT~$\pm~42\%$  scaled to 0.308~au (62~nT~$\pm~53\%$ without scaling). The percentage indicate the value of the standard deviation divided by the average. A similar reduction of the standard deviation was found in \citet{Janvier:2019}. For our sample, the reduction in the standard deviation still hold if we divide the larger of CMEs measured near 1~au into two subsamples (those in conjunction with VEX and with MESSENGER). As such, the faster expansion of more magnetized MEs on their way to 1~au has the effect of uniformizing (reducing the variance of) the ME magnetic field strength at 1~au. A 16~nT ME at 1~au may be the result of an ME with maximum magnetic field of 177~nT at 0.308~au with a fast expansion or one with a 85~nT magnetic field at 0.308~au with a slower expansion. These correspond to events 22-2012 and 23-2014 both in February 2014. This is a situation similar to the drag experienced by fast CMEs being higher than the drag experienced by slower CMEs, which result in a uniformization of the speed at 1~au, and a loss of information of what was the initial speed. While we expect the solar wind quantities to influence the CME expansion, the values measured at 1~au do not allow us to draw a conclusion regarding this. Overall, this points towards expansion being influenced by the magnetic pressure inside the ME, and therefore, some of the expansion being associated with an ``overpressure'' of the ME as compared to the background. We also find no correlation between the expansion index and the CME size near 1~au.

\begin{figure}[tb]
\centering
{\includegraphics*[width=.48\linewidth]{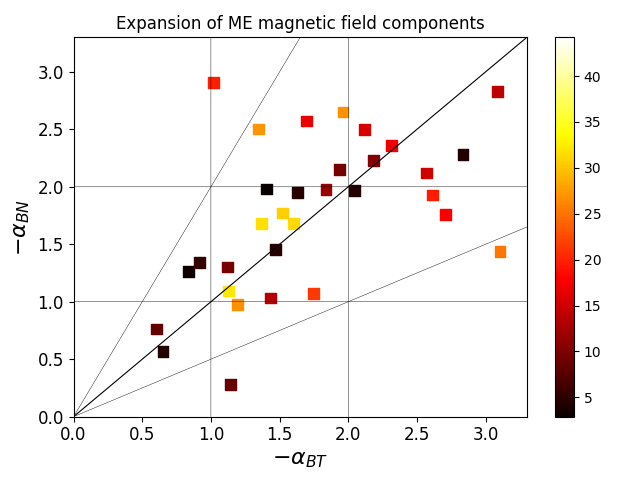}}
{\includegraphics*[width=.48\linewidth]{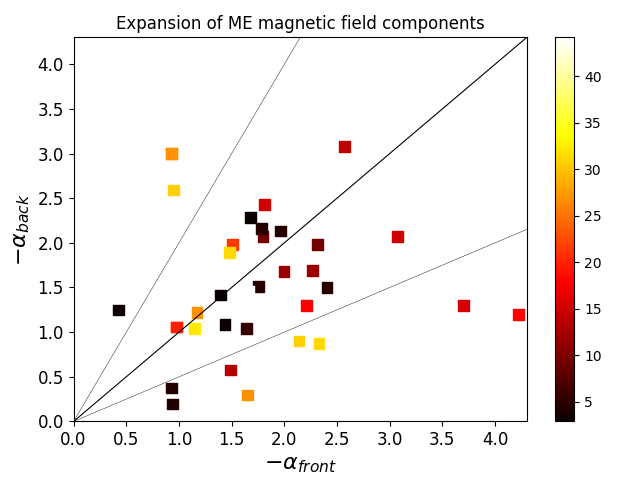}}
\caption{Left: Average of the expansion indices of the positive and negative $B_N$ component vs. average of the expansion indices of the positive and negative $B_T$ components. The lines shows the 1-to-1, 1-to-2 and 2-to-1 values as well as the expected $-1$ and $-2$ values for the indices. The colorbar is the same as in Figures 3 and 4. Right: Comparison of the expansion indices of $B_T$ and $B_N$ in the front and back of the MEs.}
\label{fig:component1}
\end{figure}

\subsection{Evolution with Distance of Magnetic Field Components Inside MEs}

\citet{Vrsnak:2019} investigated how the fitted magnetic field inside MEs and the radial size of MEs vary with distance and discussed the implications of their study for the self-similar expansion of MEs. They concluded that, for individual cases, reconnection between the ME and the solar wind and/or pancaking of the ME cross-section is necessary to understand the evolution of the ME size as compared to the evolution of the magnetic field. In a previous work,  \citet{Leitner:2007} noted that the expected difference in the decrease rate with distance of the axial and azimuthal components of the magnetic field may create differences in the trend found for the inner and outer heliosphere. \citet{Good:2019} discussed the change in orientation of the 18 CMEs measured in conjunction that they studied, finding a tendency towards lower-inclined MEs at the outer spacecraft compared to the first spacecraft. This implies a (small) difference in the way different magnetic field components change with distance.

We note that for a force-free field with self-similar expansion, the axial magnetic field is expected to vary with distance as $r^{-1}$, whereas the poloidal field should vary as r$^{-2}$. In \citet{Lugaz:2020} for the 2013 July 10-13 CME, we found that a uniform decrease of the magnetic field components as r$^{-1.6}$ was a better fit to the data than a separate fit for the $y$ (axial) or $z$ (poloidal) components of the magnetic field. Here, we continue this analysis for the 42 CMEs measured in conjunction between two spacecraft. 

Because MESSENGER and VEX were planetary missions, there are significant ``data gaps'' in the IMF measurements corresponding to the time when the spacecraft were in the planetary magnetosphere. In addition, the magnetic fields inside MEs have been reported to significantly rotate in the inner heliosphere  in some cases \citep[e.g., see][]{Nieves:2012,Winslow:2016} and it is unclear how this should be considered when comparing the axial or poloidal fields measured by these spacecraft with those measured near 1~au. As such, we compare the tangential and normal components of the magnetic field measured at the two spacecraft in the $RTN$ coordinate system. We focus on the extrema of the variation of the magnetic field components. 

For an ME that has a clear low (resp.\ high) inclination, the $B_T$ (resp.\ $B_N$) component typically keeps the same sign throughout the ME interval. In addition, the expansion of the front and back half of the ejecta may occur at different rates. For example, in event 21-2013, discussed in \citet{Lugaz:2020}, $B_T$ is always positive, while $B_N$ varies from positive to negative \citep[NWS ME following the classification of][]{Bothmer:1998}. For this event, as shown in Table~1, the $B_N$ positive component (at the front) decreases with an index of $-1.4$, whereas the $B_N$ negative (at the back) decreases with an index of $-1.1$. We therefore calculate the average of the indices for the positive and negative extrema of one component, and compare these. For the 21-2013 event, this means comparing the index of $-1.2$ for the $B_T$ unipolar component with $-1.25$ for the average of the $B_N$ indices. This shows that, although this is a low-inclined cloud, the axial and poloidal fields do not expand with a 1-to-2 ratio, but have approximately the same rate of expansion. We perform the same analysis for all MEs and these averages for the index decrease of the $B_T$ and $B_N$ components inside the ME are plotted in the left panel of Figure~\ref{fig:component1}. 

This Figure shows that there is no ME for which one component decreases as r$^{-1}$ while the other decreases as r$^{-2}$, which would be expected for the force-free expansion of a low or high inclined ME. There are a few cases for which this is approximately true. In fact for most MEs, the expansion index of the normal and tangential components agree with each other. The average of the expansion index of $B_T$ and $B_N$ are nearly identical (see Table~\ref{tab:stats} and the ratio of $\alpha_{BT}$ to $\alpha_{BN}$ is 1.09 $\pm$ 0.63. These results could occur if all MEs in our sample have an inclination close to $45^\circ$, which would imply that the normal and tangential components decrease similarly in a force-free model. This is highly unlikely; if nothing else, the four events described in Section~\ref{sec:results_spec} include MEs with a low inclination. In addition, such a situation should result in indices of both components around $-1.5$, whereas we find a cluster of MEs for which both components decrease approximately as $r^{-2}$.

Lastly, we compare the expansion of the components in the front half of the MEs with that in the back half. The results are plotted in the right panel of Figure~\ref{fig:component1}. It shows a bias towards the expansion in the front of the ejecta to be stronger than the expansion at the back. Note that we have reliable exponents only for 34 pairs (front and back) of magnetic field components, and that these are dominated by conjunctions involving MESSENGER data (28 cases vs.\ six for VEX data). The ratio of the front-to-back expansion is 1.57 $\pm$ 1.18 with 15 events with the front expansion at least 25\% larger than the back expansion and only seven for which the reverse is true (the other ten events are consistent with the same expansion in the front and the back). As this result is based on the extrema of $B_T$ and $B_N$, the exact position of the boundaries is not expected to influence the results. This is somewhat consistent with the findings of \citet{Janvier:2019} that showed that the profile of the magnetic field inside MEs is more peaked towards the front at MESSENGER and more symmetric at 1~au. This would result in the front half to show more expansion than the back half of MEs as found here. This result may be associated with the presence of a sheath region in front of the ME that allows the front part of the ME to expand relatively freely. On the contrary, the expansion of the back part of the ME may be hindered by the presence of the ME wake with speed comparable (or sometimes slightly higher) than the back of the ME. \add{The presence of fast solar wind streams behind MEs may also result in MEs being somewhat compressed in the back, and would thus still be consistent with these results. The presence of fast streams behind MEs near 1~au is a relatively frequent occurrence}. We note that this cannot be explained by aging as the back part of the ME is older than the front when it passes over a spacecraft and it is the section of the ME which has had most time to expand. This finding, if confirmed, further complicates the notion of force-free and self-similar expansion of MEs as there might not be a balance of the magnetic field at all time throughout the ME propagation.

\section{Discussion and Conclusions}  \label{sec:discuss}

\begin{figure}[tb]
\centering
{\includegraphics*[width=.98\linewidth]{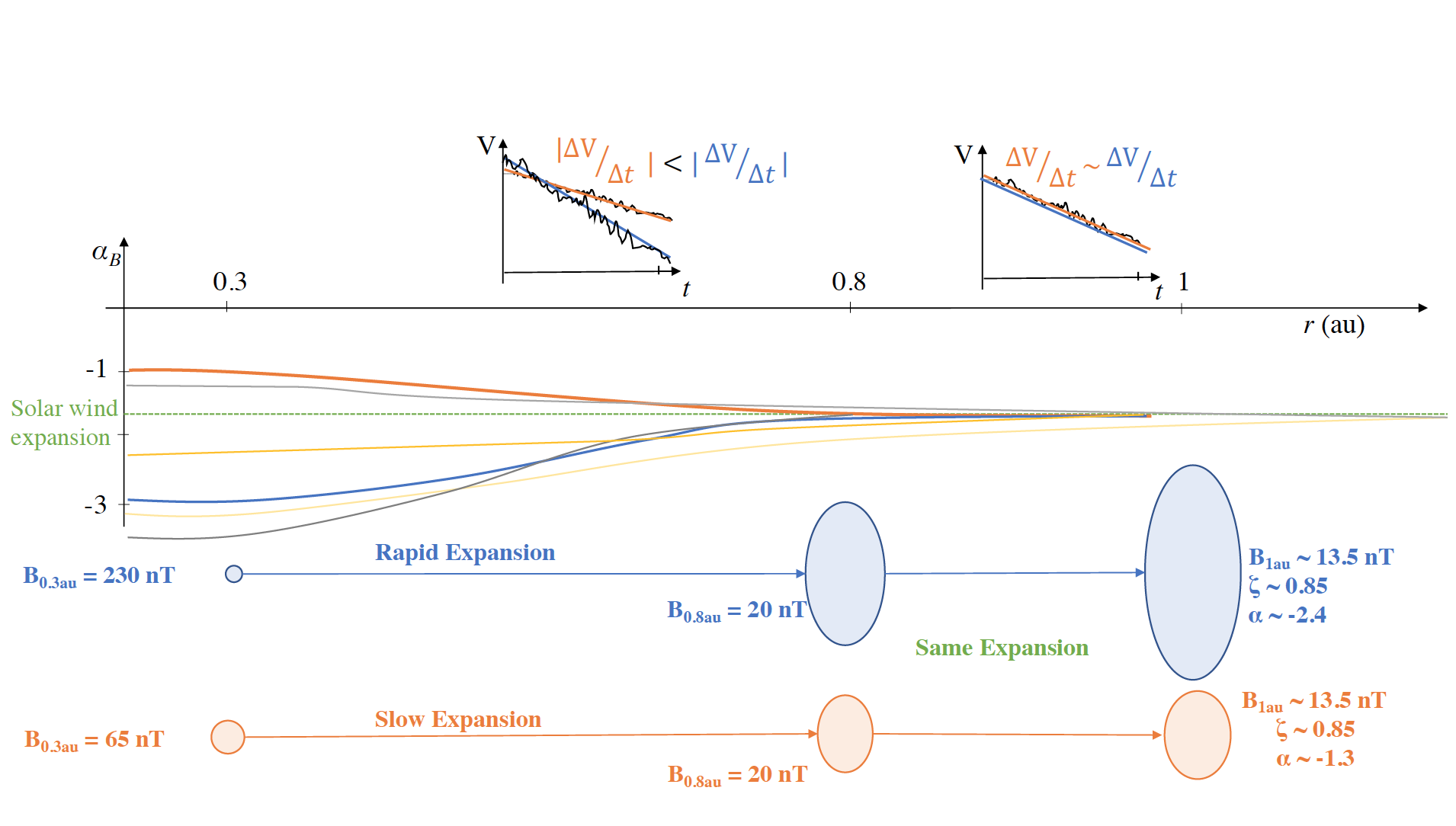}}
\caption{Schematic representation of the expansion of two MEs. In the inner heliosphere, one ME (orange) expands more slowly than the other (blue) until both reach total pressure balance with the solar wind (bottom row). Afterwards, they expand with the same rate, dictated by the solar wind expansion. Both MEs have $\alpha = -1.8$ between 0.8 and 1~au. Between 0.3 and 0.8~au, the rapidly expanding ME has $\alpha = -2.5$ and the slowly expanding ME has $\alpha = -3.2$. The combined $\alpha$ from 0.3 to 1~au are $-2.4$ and $-1.3$ for the rapidly and slowly expanding MEs, respectively. {\it In situ} measurements (top row) near 1~au do not reflect what happened in the innermost heliosphere, while measurements below $\sim 0.8$~au might reflect the conditions but are not available. Overall, the $\alpha$ of various MEs (middle row) tend towards the solar wind value (green curve) as the MEs approach 1~au with the orange and blue curves representing the MEs shown below and other colors for other potential behaviors. }
\label{fig:final}
\end{figure}

In this work, we have used {\it in situ} measurements of 42 CMEs made in conjunction by two spacecraft among MESSENGER, Venus Express, {\it Wind}, STEREO-A and STEREO-B during solar cycle 24 to compare global and local measures of ME expansion. In terms of global measures, we have focused on the index of the decrease of the magnetic field with distance, $\alpha$. In terms of local measures, we have examined the expansion speed and various dimensionless parameters, primarily $\zeta$ from \citet{Demoulin:2009} as calculated near 1~au. We have also compared the global expansion with local properties of CMEs, its initial and final speed and magnetic field strength. Our sample, in terms of average properties of the CMEs, appears typical when compared to the average properties from larger samples measured near 1~au  \citep[]{Richardson:2010,Jian:2018}. 

We have found that the global and local measures of CME expansion are, at best, only weakly correlated, indicating that measurements near 1~au do not reflect the expansion of CMEs between $\sim$ 0.3~au and 1~au. The only strong correlation has been found to occur between the ME magnetic field strength (average or maximum) at the innermost spacecraft and the index of decrease of the magnetic field. 

Overall, a picture of CME expansion in the inner heliosphere (from $\sim$ 0.3~au to 1~au) can be drawn from this work to explain the various measurements. A sketch of this scenario can be seen in Figure~\ref{fig:final}. MEs with strong internal magnetic pressure at 0.3~au expand rapidly until they reach total pressure balance with the solar wind, somewhere before 1~au (blue curves). MEs with weak internal magnetic pressure at 0.3~au expand slowly until they reach total pressure balance with the solar wind, somewhere before 1~au (orange curves). In the heliosphere near Earth ($\sim$ 0.8 to 1.1~au and probably beyond), the ME expansion is controlled by the change in the solar wind pressure \citep[]{Demoulin:2009} and it does not depend on what was the initial magnetic fiel strength. As such, both the final (near 1~au) magnetic field strength and the local measure of expansion ($\zeta$) do not reflect processes that occurred below $\sim$ 0.8~au. The power index of decrease of the magnetic field with distance, $\alpha$, is dominated by what happens in the inner heliosphere. In the example in Figure~\ref{fig:final}, both MEs have $\alpha = -1.8$ between 0.8 and 1~au for a decrease of the magnetic field by a factor of 1.5. Between 0.3 and 0.8~au, the rapidly expanding ME has $\alpha = -2.5$ for a decrease of the magnetic field by a factor of 11.6 and the slowly expanding ME has $\alpha = -3.2$ for a decrease of 3.2. The combined $\alpha$ from 0.3 to 1~au are $-2.4$ and $-1.3$, even though the ME magnetic field at 1~au is the same for both cases with a value of 13.5~nT.

While this scenario fits with the various findings in this work, to be fully tested, it would require i) more conjunction events involving three or more spacecraft, and ii) plasma measurements, especially of the velocity, in the inner heliosphere (below 0.95~au) to test the prediction that the $\zeta$ parameter may be better correlated with $\alpha$ in the innermost heliosphere.

In addition, we have found some evidence from the evolution of the tangential and normal components of the magnetic field inside MEs between the two spacecraft that MEs do not  maintain force-free conditions while they expand. This conclusion has been obtained without performing fitting of the magnetic field measurements, which would require to make assumptions regarding the morphology of the magnetic field inside MEs. In addition, fitting methods have been found to often disagree regarding ME orientation \citep[]{AlHaddad:2013}. 

Lastly, we have found evidence that the front of the ME expands faster than the back. This might be consistent with the back half of the ME being overtaken by the solar wind behind it. Such a scenario would hinder the ME expansion in its back half. This finding is consistent with the fact that many {\it in situ} measurements within MEs, such as those presented in Figure~\ref{fig:example}, have a decreasing speed profile in the front part of the ME and a constant ME speed equal to the solar wind speed in the back of the ME. This indicates that the ME expansion in the ecliptic plane is not able to continue beyond the point where the ME back speed equals the solar wind speed. This is also consistent with the lack of reverse shocks measured at the back of MEs in the ecliptic plane, contrary to what occurs at the back of stream interaction regions or MEs at high latitudes \citep[]{Gosling:1998}. 

Some of these results could be further tested if we had multi-spacecraft measurements of CMEs made at approximately the same heliocentric distance. This would allow us to compare different local measures of the CME expansion (expansion speed, $\zeta$, etc.) to determine how they vary through different crossings within the same ME. Such multi-spacecraft measurements will be possible when STEREO-A comes back to the proximity of the Sun-Earth line in 2023-2024, but this will only provide about 11 months of potential measurements within 10$^\circ$ from the Sun-Earth line. \citet{Lugaz:2018} highlighted differences between spacecraft measurements for angular separation of $\sim 0.7^\circ$; however the maximum magnetic field strength remained very consistent between two spacecraft even when the components measured by the two spacecraft showed significant differences. If such differences between MEs are common, the results about the expansion of various magnetic field components may be affected. This highlights the need for a dedicated mission providing multi-point measurements of MEs in the inner heliosphere.

\acknowledgments
This work has been made possible by the following grants: NASA NNX15AB87G, 80NSSC20K0700, 80NSSC17K0556 and 80NSSC20K0431 and NSF AGS1435785. RMW acknowledges support from NASA grant 80NSSC19K0914 and NSF grant AGS1622352. CJF acknowledges support from {\it Wind} grant 80NSSC19K1293.
All the data analyzed in this study are publicly available. MESSENGER and VEnus Express data are available on the Planetary Data System (\url{https://pds.jpl.nasa.gov}) while other data are available from the CDAWeb  (\url{https://cdaweb.sci.gsfc.nasa.gov/index.html/})

\end{document}